\documentstyle[psfig,prd,aps]{revtex}
\begin{document}
%\draft
\baselineskip 16pt
\renewcommand{\textwidth}{15cm}
\renewcommand{\topmargin}{.05cm}
\setlength{\textheight}{50pc}
\preprint{YITP-98-10,\ SU-ITP 98/04}
\tighten
\title{Chaotic New Inflation and \\ Formation of Primordial Black Holes %
%\thanks{Thank you title}
}

\author{ Jun'ichi YOKOYAMA}
\address{\hfill\\
Department of Physics, Stanford University, Stanford, CA 94305-4060 and\\
Yukawa Institute for Theoretical Physics, Kyoto University, Kyoto
606-01, Japan\\
%\hfill\\
}

\maketitle

\abstract{It is shown that in a number of scalar potentials with an unstable
local maximum at the origin chaotic inflation is followed by new inflation
if model parameters are appropriately chosen.  In this model density
fluctuation can have a large-amplitude
 peak on the comoving Hubble scale at the onset
of the slow-roll new inflation and can result in  formation of
appreciable amount of primordial black holes on astrophysically interesting
mass scales.\\

%\maketitle

%\newpage
%
%\def\theequation{\arabic{section}.\arabic{equation}}

        %%%%%%%%        Contents starts here  %%%%%%%%%%%%%%
\newcommand{\dw}{{\rm DW}}
\newcommand{\cw}{{\rm CW}}
\newcommand{\ml}{{\rm ML}}
\newcommand{\lt}{\tilde{\lambda}}
\newcommand{\lh}{\hat{\lambda}}
\newcommand{\phidot}{\dot{\phi}}
\newcommand{\phicl}{\phi_{cl}}
\newcommand{\adot}{\dot{a}}
\newcommand{\phat}{\hat{\phi}}
\newcommand{\ahat}{\hat{a}}
\newcommand{\hhat}{\hat{h}}
\newcommand{\phihat}{\hat{\phi}}
\newcommand{\Nhat}{\hat{N}}
\newcommand{\hth}{h_{th}}
\newcommand{\hbh}{h_{bh}}
\newcommand{\gsim}{~\mbox{\raisebox{-1.0ex}{$\stackrel{\textstyle >}
{\textstyle \sim}$ }}}
\newcommand{\lsim}{~\mbox{\raisebox{-1.0ex}{$\stackrel{\textstyle <}
{\textstyle \sim}$ }}}
\newcommand{\bfx}{{\bf x}}
\newcommand{\bfy}{{\bf y}}
\newcommand{\bfr}{{\bf r}}
\newcommand{\bfk}{{\bf k}}
\newcommand{\bkp}{{\bf k'}}
\newcommand{\order}{{\cal O}}
\newcommand{\beq}{\begin{equation}}
\newcommand{\eeq}{\end{equation}}
\newcommand{\beqa}{\begin{eqnarray}}
\newcommand{\eeqa}{\end{eqnarray}}
\newcommand{\mpl}{M_{Pl}}
\newcommand{\lmk}{\left(}
\newcommand{\rmk}{\right)}
\newcommand{\lkk}{\left[}
\newcommand{\rkk}{\right]}
\newcommand{\lnk}{\left\{}
\newcommand{\rnk}{\right\}}
\newcommand{\call}{{\cal L}}
\newcommand{\calr}{{\cal R}}
\newcommand{\half}{\frac{1}{2}}
\newcommand{\kc}{\kappa\chi}
\newcommand{\bkc}{\beta\kappa\chi}
\newcommand{\gkc}{\gamma\kappa\chi}
\newcommand{\gbkc}{(\gamma-\beta)\kappa\chi}
\newcommand{\dchi}{\delta\chi}
\newcommand{\dphi}{\delta\phi}
\newcommand{\dOmega}{\delta\Omega}
\newcommand{\Phibd}{\Phi_{\rm BD}}
\newcommand{\echi}{\epsilon_\chi}
\newcommand{\ephi}{\epsilon_\phi}
\newcommand{\Phihat}{\hat{\Phi}}
\newcommand{\Psihat}{\hat{\Psi}}
\newcommand{\that}{\hat{t}}
\newcommand{\Hhat}{\hat{H}}
\newcommand{\zk}{z_k}
\newcommand{\msolar}{M_\odot}

%\baselineskip 0.8cm
%\newpage
\section{Introduction}

If overdensity of order of unity exists in the hot early universe,
a black hole can be formed when the perturbed region enters the
Hubble radius.  While the properties of the primordial black holes
(hereafter PBHs) thus produced were a subject of extensive study
decades ago \cite{Zel,Haw}, there were no observational evidence of
their existence and only
observational constraints were obtained against their
mass spectrum \cite{Carr,Nov}.
Recently, however, possibilities of their existence have been
raised from a number of astrophysical and cosmological considerations
and there is an increasing interest in it.

For example, they may be
the origin of the massive compact halo objects (MACHOs) which are
dark compact objects with typical mass $\sim 1\msolar$ and make up about
$\order (10^{-2})$ of the critical density \cite{macho}.
While the primary MACHO candidates are substellar baryonic objects such
as brown dwarfs, it is difficult to reconcile such a large amount of
these objects with the observed mass function of low mass stars
\cite{RF} and with the infrared observation of dwarf component \cite{BU},
unless the mass function is extrapolated to the lower masses in an
extremely peculiar manner of Population III stars are produced abundantly
at the relevant mass scale \cite{pop3}.  Hence we should consider PBHs
seriously as the second option.  Furthermore this possibility can be
experimentally tested by observing gravitational radiation from
coalescing black hole MACHO binaries by laser interferometers \cite{Naka}.

Another interesting possibility is PBHs with mass $M\sim 10^{15}$g
which are just evaporating now \cite{evaporate}.  It has been argued that
high energy phenomenon associated with evaporation can explain origin of a
class of gamma-ray burst.  For this purpose their  abundance
 should be around $\Omega=10^{-8}$
in unit of the critical density \cite{Cline}.

One may also consider formation of much heavier black holes with mass
$M\sim 10^8\msolar$.  Such black holes are expected to exist in the
center of AGNs and quasars \cite{Turner}
and act as an engine of their activity.
Although it is generally
believed that these black holes are formed after recombination
with a specially arranged spectrum of density fluctuations
\cite{Loeb}, it might be interesting to consider the possibility that
they were of primordial origin, which might lead to a
new scenario of galaxy formation.

In any case, in order to produce PBHs on some specific scale, we
must prepare density perturbation whose amplitude has a
high peak of $\order (10^{-2})$ on the corresponding scale.  It is
difficult, however, to realize such a spectral shape in inflationary
cosmology \cite{oriinf,newinf,chaoinf,inf}, which is not only
indispensable to solve the horizon and the flatness problems but
also the most sensible way to generate density perturbations
\cite{pert}, because usual models predict a scale-invariant
spectrum.  Given the smallness of the observed anisotropy of the
cosmic microwave background radiation (CMB) \cite{COBE}, the
amplitude of primordial density fluctuation turns out to be
$\order (10^{-5})$ on any observable  scales in most common
models.

It is possible, nevertheless, to realize a non scale-invariant
spectrum by choosing somewhat contrived  forms of the inflaton's
potential.  While examples of the various perturbation spectrum
realized in different  potentials  is found in \cite{Hodges}, we
must admit that few of the models giving non scale-invariant
spectrum with a single scalar field has a motivation in sensible
particle physics.  In particular, the model proposed by Ivanov
et al.\ \cite{INN} in order to produce significant amount of
PBHs employs a scalar potential with two breaks and a plateau in
between.
Several other toy potentials  have also been studied in
\cite{BP}.

One can construct more natural inflation models for PBH formation
if one allows multiple scalar fields.  Some of them make use of
primordially isocurvature fluctuations \cite{JY}, others include
multiple stages of inflation with each regime governed by different
field \cite{ST,Ra,GL,Kawa}.
We must assign appropriate form of coupling of the scalar fields
in each model, which may not always be easy.

\newpage

In the present paper, we propose a new scenario of multiple
inflation which, unlike previous double inflation models
\cite{ST,Ra,GL,Kawa,di},
contains only one source of inflation.  In this model we employ in the
Einstein gravity an
inflaton scalar field, $\phi$, with a
simple potential, $V[\phi]$, which has an unstable local maximum
at the origin, such as a quartic double-well potential.
This is the same setup as the new inflation scenario \cite{newinf},
but chaotic inflation is also possible if the $\phi$ has a sufficiently
large amplitude initially.  In fact Brandenberger and Kung \cite{BK}
studied the initial distribution of a scalar field with such a
potential and concluded that chaotic inflation is much more likely than new
inflation.  Thus we also start with chaotic inflation, but  show that
new inflation is also possible when $\phi$ evolves towards the origin
after chaotic inflation if the parameters of the potential is
appropriately chosen so that the scalar field has the right amount of
kinetic energy after chaotic inflation in order to climb up the
potential hill  near to the origin and start slow rollover there.
Hence in this model the initial condition for new inflation is realized
not due to the high-temperature symmetry restoration nor for a topological
reason \cite{topological}, but by dynamical evolution of
the field which has already become sufficiently
homogeneous because of the first stage
of chaotic inflation.  We shall refer this succession of inflation
simply to chaotic new inflation.

With an appropriate shape of the potential, density
fluctuations generated during new inflation can have larger amplitude
than those during chaotic inflation.  Furthermore, since the power
spectrum of fluctuation generated during new inflation can be tilted,
it can have a peak on the comoving Hubble scale when the inflaton enters
the slow-rollover phase during new inflation.  If the peak amplitude is
sufficiently large, it results in formation of PBHs on the horizon mass
scale when the corresponding comoving scale reenters the Hubble radius
during radiation domination.  We shall show below that such a scenario
is indeed possible with a simple shape of the inflaton's potential.

The rest of the paper is organized as follows.  In \S II we summarize
basic features of chaotic inflation and new inflation scenarios and
in \S III we calculate formation probability of PBHs in our model.
Then we report the results of numerical analysis with various potentials
in \S IV and \S V.  Finally \S VI is devoted to conclusion.

\section{Chaotic Inflation and New Inflation}

First let us consider two potentials which have a global minimum at
$\phi=v < \mpl$:
\beqa
 V_\cw[\phi]&=&\frac{\lt}{4}\phi^4\lmk \ln\left|\frac{\phi}{v}\right|
  -\frac{1}{4}\rmk+\frac{\lt}{16}v^4,  \label{cw} \\
 V_\dw[\phi]&=&\frac{\lambda}{4}(\phi^2-v^2)^2. \label{dw}
\eeqa
The former is the
Coleman-Weinberg potential \cite{CW},
which was employed in the original version of the new
inflation scenario \cite{newinf},
and the latter is a simple double-well potential.
For $\phi \gg \mpl > v$, the both potentials
can be approximated at least locally by a simple quartic potential
$V[\phi]=\frac{\lambda}{4}\phi^4$ with $\lambda \simeq
\lt\ln\left|\phi/v\right|$ for the former case, and the
evolution of the universe in this regime is practically the same as that in
chaotic inflation with the quartic potential \cite{chaoinf}.
That is, from the
slow-rollover equations of motion,
\beqa
    3H\phidot &=& -V'[\phi],  \\
  H^2=\lmk\frac{\adot}{a}\rmk^2&=&\frac{8\pi}{3\mpl^2}V[\phi],
\eeqa
we find
\beqa
   \phi (t)&=&\phi_i\exp\lmk -\sqrt{\frac{\lambda}{6\pi}}\mpl t\rmk, \\
  a(t)&=& a_i\exp\lkk\frac{\pi}{\mpl^2}\lmk\phi^2_i-\phi^2(t)\rmk\rkk,
\eeqa
which are valid until the slow-rollover conditions
$|\dot{H}| \ll H^2,$ $\ddot{\phi} \ll 3H\phidot$ etc.\ break down at
around $\phi_e \simeq \max\lmk \frac{\mpl}{8\pi}, 4v\rmk$.
Here the suffix $i$ implies the initial value at $t=t_i$.
The $e$-folding number of (chaotic) inflationary expansion between $\phi$
and $\phi_e$, $N(\phi\rightarrow\phi_e)$, is therefore given by
\beq
  N(\phi\rightarrow\phi_e)=\frac{\pi}{\mpl^2}\lmk\phi^2-\phi^2_e \rmk
 \simeq \frac{\pi}{\mpl^2}\phi^2,
\eeq
the latter approximate equality being valid for $\phi^2 \gg \phi_e^2$.

The amplitude of linear curvature perturbation, $\Phi$,
generated on the comoving scale,
$l=2\pi/k$,  is given by
\beq
  \Phi\lmk l=\frac{2\pi}{k}\rmk
 = \left. f\frac{H}{|\phidot|}\dphi\right|_{t_k}
=\left.f\frac{H^2}{2\pi|\phidot|}\right|_{t_k} \equiv f\Delta (t_k).
\eeq
Here $t_k$ is the epoch when the wavenumber $k$  satisfied
$k=a(t_k)H(t_k)$ during inflation \cite{pert},
and
$\dphi=H(t_k)/(2\pi)$ is the root-mean-square amplitude  of fluctuation
accumulated during one expansion time around $t=t_k$, and
$f=3/5$ $(2/3)$ in the matter- (radiation-) dominated stage.
The above expression is valid until the
comoving scale $l$ crosses the Hubble radius again.
In the case with quartic inflaton potential, it reads
\beq
  \Phi(l)=f\sqrt{\frac{2\pi\lambda}{3}}\lmk\frac{\phi(t_k)}{\mpl}\rmk^3
\simeq \frac{f}{\pi}\sqrt{\frac{2\lambda}{3}}
N_{ke}^{\frac{3}{2}},
\eeq
with $N_{ke}\equiv N(\phi(t_k)\rightarrow\phi_e)$.
The large-angular-scale anisotropy of CMB
  due to the Sachs-Wolfe effect \cite{SW} is then given by
\beq
  \frac{\delta T}{T}=\frac{1}{3}\Phi= \left.\frac{H^2}{10\pi|\phidot|}
\right|_{t_k}.
\eeq
In the usual case with single stage of inflation one typically takes
$N_{ke}\simeq 60$ and determine the value of $\lambda$ using the COBE
data $\delta T/T \simeq 1\times 10^{-5}$ to yield
$\lambda \simeq 2\times 10^{-13}$ \cite{Salopek}.
In the case with second stage of
inflation the value of $N_{ke}$ corresponding to the scale observed by COBE
can be significantly smaller and the normalized
value of $\lambda$ can be different.
Taking $N_{ke}=20$ with another 40 $e$-folds during new inflation,
for example, we find $\lambda \simeq 6\times 10^{-12}$.

After chaotic inflation, the scalar field starts oscillation around
$\phi=v$ if $v$ is too large, or it overshoots the symmetric state
$\phi=0$ and approaches the other minimum $\phi=-v$ if $v$ is too small.
If $v$ is appropriately chosen at an intermediate value,
on the other hand, it can spend a long enough time
near the origin and then slow-rollover new inflation can set in,
which ends up with either $\phi=v$ or $\phi=-v$ depending on the sign
of $\phi$ when classical slow-rollover regime begins.
We must resort to a precise
numerical integration of equations of motion to find out what values
of $v$ to take.

Before doing
this,  let us review the basic properties of new inflation, assuming
that the slow-rollover starts from $\phi=\phi_s$ at the epoch $t=t_s$.
Since the shape of the potential around the origin is different between
the Coleman-Weinberg potential and the double-well potential, we must
treat them separately.

First for the former, the potential near the origin can be approximated
by
\beq
 V_\cw[\phi]\simeq V_0 -\frac{\lh}{4}\phi^4,
\eeq
with $\lh=\lt\ln|\phi_s/v|$ and $V_0$ reads $V_0=\lt v^4/16$.
The classical slow-rollover solution satisfies
\beqa
    H^2&=&H^2_0=\frac{8\pi V_0}{3\mpl^2},\\
  \lh \phi^2 &=& \frac{3H_0^2}{2N(\phi\rightarrow\phi_f)+1}
 \simeq \frac{3H_0^2}{2N(\phi\rightarrow\phi_f)},  \label{cwclas}
\eeqa
the latter approximate equality being valid for $N \gg 1$.
  The curvature perturbation reads
\beq
  \Phi(l)=\frac{f}{\pi}\sqrt{\frac{2\lh}{3}}
  N_{kf}^{\frac{3}{2}},
\eeq
where $N_{kf}$ is defined by
$ N_{kf}\equiv N(\phi(t_k)\rightarrow \phi_f)$.
Since $\lh$ is no larger than $\sim 20\lambda \sim 10^{-10}$
the resultant curvature
perturbation is at most of order of $10^{-4}$ for $N_{kf}<60$.  Hence
chaotic new inflation with the Coleman-Weinberg potential is not suitable
to produce large enough density perturbation on small scale to
warrant significant formation of primordial black holes.

Next we consider new inflation with the double-well potential (\ref{dw}),
which can be  approximated as
\beq
 V_\dw[\phi]\simeq V_0 -\frac{1}{2}m^2\phi^2,   \label{apote}
\eeq
with $V_0=\frac{\lambda}{4}v^4$ and $m^2=\lambda v^2$, but for later
convenience we treat as if $V_0$ and $m^2$ were free parameters.
The classical slow roll-over solution reads
\beq
  \phi(t) = \phi_s\exp\lkk \frac{m^2}{3H_0^2}H_0(t-t_s)\rkk,~~~
H_0^2= \frac{8\pi V_0}{3\mpl^2},~~~~a(t)\sim e^{H_0t}.
\label{dwclas}
\eeq
Inflation ends at $\phi=\phi_f\equiv cv$ with $c$ being a constant of
order of 0.1, so we find
\beq
  N(\phi\rightarrow \phi_f) = \frac{3H_0^2}{m^2}\ln\frac{\phi_f}{\phi}
  =  \frac{3H_0^2}{m^2}\ln\frac{cv}{\phi},
\eeq
and curvature perturbation  in the slow-rollover regime reads
\beq
  \Phi(l)=f\Delta_{\rm SR}(t_k)\equiv
  3f\frac{H_0^2}{m^2}\frac{H_0}{2\pi\phi(t_k)}=
 \frac{3f}{2\pi}\frac{H_0^3}{m^2\phi_f}
 \exp\lmk \frac{m^2}{3H_0^2}N_{kf}\rmk.  \label{dwcurv}
\eeq

The above classical slow-roll solution (\ref{dwclas}) (and also
(\ref{cwclas}) for the Coleman-Weinberg model) is valid only when
classical potential force dominates diffusion due to quantum fluctuation.
In other words, the classical motion during
one expansion time, $\Delta\phi=|\dot{\phi}|H^{-1}$, should be
larger than the amplitude of quantum fluctuation,
$\delta\phi=H/(2\pi)$, accumulated in the same period.  For the
potential (\ref{apote}) this condition reads
\beq
   |\phi| > \frac{3H_0^3}{2\pi m^2}\equiv \phi_q.  \label{SR}
\eeq
If $\phi$ stays in the region $|\phi|<\phi_q$ for more than
one expansion time, quantum fluctuation will start to dominate the
dynamics and the universe enters a self-reproduction regime
which can last eternally \cite{eternal}.  Then the memory of
chaotic inflation would be washed away and our double inflation
scenario would not work.

Note that  the
amplitude of density fluctuation evaluated with the slow-roll
approximation  exceeds unity, $\Delta_{\rm SR}(t_k)>1$, for $|\phi|<\phi_q$.
This means that if $\phi$ entered the slow-roll regime in
$|\phi|<\phi_q$, we would find the peak amplitude of density
fluctuation larger than unity.  Since our goal is to produce a
spectrum of density fluctuation whose amplitude at the peak is
at most $\order (10^{-2})$, we should arrange model parameters so
that $\phi$ crosses over  the
region $|\phi|<\phi_q$ rapidly enough to ensure
$\Delta(t_k) \ll \Delta_{\rm SR}(t_k)$ in this period and to
avoid self-reproduction, or $\phi$ should reverse its direction of
motion before reaching $\phi=\phi_q$.  In either case, as long as we
choose the model parameters so that the peak amplitude of
fluctuation is $\order (10^{-2})$, we can automatically avoid the
dangerous self-reproduction in the new inflation regime.

Equation (\ref{dwcurv}) implies the spectral index
of power-law density fluctuation
is given by
\beq
  n=1-\frac{2m^2}{3H_0^2} <1,
\eeq
Thus the power spectrum can be significantly tilted.
The linear perturbation has a peak amplitude
\beq
  \max\Phi=3f\frac{H_0^2}{m^2}\frac{H_0}{2\pi\phi_s},
\eeq
on the comoving scale $l_s\equiv 2\pi/k_s$ where $k_s\equiv a(t_s)H(t_s)$.
It can be large if $\phi_s$ turns out to be small.
Since the coupling constant is determined by COBE observation,  we have
essentially one parameter, $v$, which determines not only the ratio
$H^2_0/m^2$ but also $\phi_s$.  Hence we must await the result of
numerical solution of the equations of motion in order to see if
this model realizes large enough density perturbation on cosmologically
interesting scales.

Before proceeding to the numerical analysis, we next calculate the
abundance, if at all, of the PBHs produced due to the potentially large
fluctuations generated in the new inflation regime with the potential
(\ref{apote}).

\section{Abundance of Primordial Black Holes}

The initial fraction of the PBHs with mass $M$, $\beta(M)$,
produced during radiation
dominated era has been estimated by Carr \cite{Carr} as
\beq
  \beta(M)\simeq \delta(M)\exp\lmk -\frac{1}{18\delta^2(M)}\rmk,
  \label{Gauss}
\eeq
assuming the density fluctuation is Gaussian distributed with the
root-mean-square at the horizon crossing given by $\delta(M)$.
Since PBHs are formed at high density peaks with $\delta \gsim 1/3$,
however, possible non-Gaussian effect on the tail may significantly
affect the estimation of their abundance, as was first pointed out by
Bullock and Primack \cite{BP}.  Here we estimate the abundance of
PBHs produced in our model by calculating the probability distribution
function (PDF) of the coarse-grained field in terms of the stochastic
inflation approach\cite{Staro},
following more recent analysis by Ivanov \cite{Ivanov}.

Since our model predicts a relatively sharp peak at the scale $l_s$
 corresponding
to the onset of slow-rollover
we estimate the abundance of PBHs on this particular scale with the
potential
\beq
  V[\phi]=V_0 - \frac{1}{2}m^2\phi^2.
\eeq
The scalar field coarse-grained over a super-Hubble scale,
$\phat (t,\bfx)$ satisfies the Langevin equation
\beq
  3H_0\dot{\phat}+V'[\phat]=q(t,\bfx), \label{Langevin}
\eeq
with the temporal correlation function of the stochastic force,
$q(t,\bfx)$, given by
\beq
  \langle q(t,\bfx)q(t',\bfx) \rangle = \frac{9H_0^5}{4\pi^2}\delta(t-t').
\eeq

Since the potential is a quadratic function,
the PDF of $\phat$, $\Pi[\phat=\phi,t]$ remains Gaussian
as long as the initial distribution is also Gaussian.
For example, starting with the PDF $\Pi [\phi,t_s]=\delta(\phi-\phi_s)$
and solving the Fokker-Planck equation
\beq
\frac{\partial \Pi}{\partial t}=\frac{1}{3H_0}
\frac{\partial ~}{\partial\phi}V'[\phi]\Pi + \frac{H_0^3}{8\pi^2}
\frac{\partial^2\Pi}{\partial\phi^2} \equiv
-\frac{\partial J}{\partial\phi}, \label{FP}
\eeq
with the boundary condition $\Pi[\phi\rightarrow\pm\infty,t]=0$,
we find
\beq
 \Pi[\phi,\tau]=\frac{1}{\sqrt{2\pi}\sigma(\tau)}\exp\lmk -
 \frac{(\phi-\phicl(\tau))^2}{2\sigma^2(\tau)}\rmk,~~~\tau\equiv t-t_s,
   \label{solution}
\eeq
with
\beqa
  \phicl(\tau)&=&\phi_s\exp\lmk\frac{m^2}{3H_0^2}H_0\tau\rmk,  \\
  \sigma^2(\tau)&=&\frac{3H_0^4}{8\pi^2m^2}\lkk
  \exp\lmk\frac{2m^2}{3H_0^2}H_0\tau\rmk-1\rkk.
\eeqa

This does not
imply, however, that the resultant abundance of PBHs are given in the
form (\ref{Gauss}), because of the nonlinear dependence of the
metric perturbations.  Following Ivanov \cite{Ivanov}, we write
the coarse-grained metric in the quasi-isotropic form
\beq
  ds^2=-dt^2+\ahat^2\lmk\phat(t,\bfx)\rmk d\bfx^2,
\eeq
where the scale factor $\ahat$ now depends on the coarse-grained
spatial coordinate as well, and quantify the metric perturbation in terms
of
\beq
  \hhat \equiv \frac{\ahat(t,\bfx)}{a(t)}-1,
\eeq
which is frozen in the super-Hubble scales.
In the limit $\hhat \ll 1$, $\hhat$ reduces to the gauge-invariant growing
adiabatic perturbation \cite{KS}.

We are interested in the statistical distribution of metric perturbation
on the comoving scales that leave the Hubble radius in the period
between $\tau=0$ and $\tau\simeq H_0^{-1}$ when the classical solution
has rolled down to $\phicl (\tau=H_0^{-1})=
\phi_s\exp\lmk\frac{m^2}{3H_0^2}\rmk\equiv \phi_{s1}$.
While the amplitude of the metric perturbation is determined by the
duration of inflation in each region, it is not straight forward to
extract perturbation on these particular scales in the present
approach in which nonlinear effects are at least partially
taken into account.
In this situation, in order to obtain the PDF for the duration of
inflation in a specific regime, Ivanov \cite{Ivanov} proposed
to solve the Fokker-Planck equation (\ref{FP}) with the initial
condition $\Pi[\phi,0]=\delta(\phi-\phi_s)$ and the absorbtive
boundary condition at $\phi=\phicl(\tau=H_0^{-1})$, and to identify
the probability current $J$ in (\ref{FP}) with the desired PDF,
from which he finds the PDF of $\hhat$, $P[\hhat]$, as
\beq
  P[\hhat]=\frac{1}{\sqrt{2\pi\Delta^2\Nhat}}\frac{N_{cl}}{\Nhat}
 \exp\lmk -\frac{(\hat{N}-N_{cl})^2}{2\Delta^2 \hat{N}}\rmk
  \frac{d\hat{N}}{d\hhat},
  \label{Ivh}
\eeq
in the case with a linear potential, where $\Nhat$ is the $e$-folding
number of expansion while the classical solution allows the $e$-folding
number of $N_{cl}$, to which we  put unity here. $\hhat$ and
$\hat{N}$ are related by
\beq
  \hhat=\exp(\hat{N}-N_{cl})-1.
\eeq

Since our potential is more complicated, it is cumbersome to adopt the
same procedure.  So we instead use the solution (\ref{solution})
and obtain the PDF of the duration of inflation while
$\phihat$ evolves from $\phi_s$ to $\phi_{s1}$, $\hat{\tau}$,
simply assuming that
$\phihat$ satisfies the classical equation of motion at the moment
$\phihat=\phi_{s1}$.  Then we find the PDF of $\hhat$ as
\beq
  P[\hhat=h]=\frac{m^2}{3H_0^2}\frac{\phi_{s1}}{1+h}\Pi\lkk \phi_{s1},
  \frac{1}{H_0}\lmk 1+\ln(1+h)\rmk\rkk.  \label{msol}
\eeq

Before proceeding to the calculation of PBH abundance, we mention
the relation between this approach and Ivanov's result (\ref{Ivh}).
If we follow the
same procedure as above for a linear potential we obtain explicitly
\beq
    P[\hhat]=\frac{1}{\sqrt{2\pi\Delta^2\Nhat}}
 \exp\lmk -\frac{(\hat{N}-N_{cl})^2}{2\Delta^2 \hat{N}}\rmk
  \frac{d\hat{N}}{dh}.
\eeq
Thus the only difference to (\ref{Ivh}) is the absence of a prefactor
$N_{cl}/\hat{N}$.  One may also be tempted to define the PDF for
$\hat{\tau}$ from the probability current $J$ using a solution analogous to
 (\ref{solution}) but with a linear potential.
In this case one would find
\beq
   P[\hhat]=\frac{1}{\sqrt{2\pi\Delta^2\Nhat}}
  \lmk\frac{1}{2}+\frac{N_{cl}}{2\hat{N}}\rmk
  \exp\lmk -\frac{(\hat{N}-N_{cl})^2}{2\Delta^2 \hat{N}}\rmk
  \frac{d\hat{N}}{dh}.
\eeq
Again we only have an extra prefactor of order of unity.
Since the deviation between $\hat{N}$ and $N_{cl}$
is not expected to be too large in our case, unlike in the situation
of eternally self-reproducing inflationary universe \cite{eternal},
all the above
three approach will yield essentially the same result as far as the rare
PBH formation is concerned.  This also implies that our semi-nonlinear
approach suffices to the present problem.

We now proceed to evaluation of the fraction of PBHs produced using
(\ref{msol}), which is explicitly written as
\beq
  P[h]=\frac{\gamma (1+h)^{-1}}{\sqrt{2\pi c^2}
\lkk (1+h)^{2\gamma} - e^{-2\gamma}\rkk^{1/2}}\exp\lnk -
\frac{\lkk (1+h)^\gamma -1\rkk^2}{2c^2 \lkk (1+h)^{2\gamma}
- e^{-2\gamma}\rkk}\rnk,  \label{hprob}
\eeq
with
\beq
   c^2\equiv\frac{3H_0^4}{8\pi^2m^2\phi_s^2},~~{\rm and}~~
   \gamma\equiv\frac{m^2}{3H_0^2}.
\eeq
Since $\phi$ is slowly rolling at $\phi=\phi_s$ one can also write
$c^2$ in terms of $\Delta$ as
\beq
  c^2=\frac{\gamma}{2}\Delta^2(t_s)\equiv \frac{\gamma}{2}\Delta_s^2.
  \label{cgam}
\eeq

In the limit of small $h$ it properly reduces to Gaussian:
\beq
   P[h]\longrightarrow \frac{\gamma}{\sqrt{2\pi c^2(1-e^{-2\gamma})}}
 \exp\lkk - \frac{\gamma^2 h^2}{2c^2(1-e^{-2\gamma})}\rkk.
\eeq
If we further take the limit $\gamma\longrightarrow 0$, it reduces to
a more familiar form:
\beq
  P[h]\longrightarrow \frac{1}{\sqrt{2\pi\Delta_s^2}}
  e^{-\frac{h^2}{2\Delta_s^2}}.
\eeq

From (\ref{hprob}) the probability that $h$ exceeds a threshold value
$\hth$ is estimated as
\beqa
  P[h>\hth]&=&\int_{\hth}^\infty P[h]dh \nonumber \\
 &\simeq& \frac{c}{\sqrt{2\pi}}
\frac{\lkk (1+\hth)^{2\gamma}-e^{-2\gamma}\rkk^{3/2}}
{(1+\hth)^\gamma\lkk (1+\hth)^\gamma-1\rkk
 \lkk (1+\hth)^\gamma -e^{-2\gamma}\rkk}  \label{hthprob} \\
&~&~~~~~~\times\exp\lnk-\frac{\lkk (1+\hth)^\gamma-1\rkk^2}
{2c^2\lkk (1+\hth)^{2\gamma}- e^{-2\gamma}\rkk}\rnk, \nonumber
\eeqa
which is insensitive to the upper end of integral since the dominant
contribution comes from $h\simeq \hth$.
The criterion for black hole formation has been numerically investigated
by Nadegin, Novikov, and Polnarev \cite{NNP} and by Biknell and Henriksen
\cite{BH}.  Although it depends on the shape of the perturbed region,
the generic value of the threshold reads $\hth > 0.75-0.9.$  We take
the black hole threshold as $\hbh=0.75.$

Putting $\hth=\hbh$ we can identify (\ref{hthprob}) with the fraction
of primordial black holes,
\beq
  \beta(M_s)=P[h>\hbh],
\eeq
where the typical black hole mass, $M_s$, is equal to the horizon mass
when the comoving scale $l_s$ reenters the Hubble radius.
Note that since
our model predicts density fluctuation which is highly peaked
on the comoving scale $l_s$ the resultant mass function of PBHs are
sharply peaked at the mass around $M_s$.

\section{Numerical analysis of  chaotic new inflation with the simplest
potentials}

We now proceed to the numerical analysis of the equations of motion for
the homogeneous part;
\beqa
   \ddot{\phi}(t)+3H\phidot+V'[\phi]=0,  \label{phieq}\\
   \lmk\frac{\dot{a}}{a}\rmk^2=
H^2(t)=\frac{8\pi}{3\mpl^2}\lmk \frac{\phidot^2}{2}+V[\phi]\rmk,
\label{aeq}
\eeqa
with either $V[\phi]=V_\cw[\phi]$ or $V[\phi]=V_\dw[\phi]$.
One of the nice features of chaotic inflation is that no matter what
initial configuration is adopted, the system rapidly approaches the
slow-rollover solution with exponential accuracy once inflation sets in
with sufficiently large initial value of $\phi$.  We stress in
this sense
that the dynamics of the second inflationary phase is independent of the
choice of initial condition but is only sensitive to the model parameters
of the potential, and unlike in the original version of new inflation
fine-tuning of initial condition is not necessary.  For our purpose, it
is sufficient to take $\phi \gsim 3.5\mpl-4\mpl$.

First, for completeness we solve equations (\ref{phieq}) and (\ref{aeq})
 for new inflation with the Coleman-Weinberg potential (\ref{cw}).
We find that if $0.2201\mpl < v < 0.2259\mpl$
at least ten $e$-folds inflationary
(accelerated) expansion with $|\dot{H}| < H^2$ is realized independent
of the value of $\lt$.  For $0.2223\mpl < v < 0.2239\mpl$
we have more than 60
$e$-folds new inflation and in this case the chaotic inflation
regime would be inflated beyond our current Hubble radius.
In any case, as we mentioned in \S II we cannot obtain large-amplitude
 fluctuations in the observable scale in this model.

Next we report the case of the double-well potential (\ref{dw}).
In this case we find,  independent of the value of $\lambda$, that
the field settles down to the positive minimum if $v \geq 0.16286751\mpl$
and it overshoots the origin to a negative value for
$v \leq 0.16286750\mpl \equiv v_{cr}$.
If $v$ is much smaller than $v_{cr}$,
$\phi$ will travel between positive and negative values several
times before settling down to one of the minima, but we do not
consider this possibility here.

For $v \simeq v_{cr}$, we find the vacuum energy at the origin
induces inflationary expansion which lasts only about five $e$-folds.
This is because the curvature of the potential at the origin is
so large that the universe cannot stay in the slow-roll phase as
is seen from the fact that the ratio
\beq
    \frac{m^2}{3H_0^2}=\frac{1}{2\pi}\lmk\frac{\mpl}{v_{cr}}\rmk^2
    \simeq 6.00,
\eeq
is substantially larger than unity.
As a result the magnitude of density fluctuations produced in this
regime remains small: $\Delta=\order (10^{-5})$ with COBE-normalized
self coupling $\lambda=3\times 10^{-13}$.

We thus find neither the Coleman-Weinberg potential (\ref{cw})
nor the double-well potential (\ref{dw}) lead to formation of
large-amplitude density fluctuations on currently observable
scales in the chaotic new inflation scenario.  But the reasons of
their failures  are opposite; with the Coleman-Weinberg potential,
the field moves too slowly and the scales with large-amplitude
fluctuations are inflated away, while with the double-well
potential, $\phi$ moves so rapidly that $\Delta$ remains too small.

\section{PBH formation in chaotic new inflation}

The above observation naturally leads us to consider a different class
of potential including another free parameter:
\beq
  V_\ml [\phi]=-\frac{1}{2}m^2\phi^2+
  \frac{\lt}{4}\phi^4\lmk \ln\left|\frac{\phi}{v}\right|
  -\frac{1}{4}\rmk + V_0,  \label{ML}
\eeq
that is, typical one-loop effective potential with nonvanishing
mass term at the origin.  This type of the potential with a
positive mass-squared at the origin was employed in the original
inflation scenario \cite{oriinf}, but for our purpose we adopt a
negative mass term.

The potential (\ref{ML}) has four parameters, but one of them, $V_0$
is fixed from the requirement that the vacuum energy density
vanishes at the potential minimum $\phi\equiv\pm\phi_m$.
While we numerically solve the equation $V'[\phi_m]=0$ and
obtain the numerical value of  $V_0$ so that $V[\phi_m]=0$,
we can also calculate them perturbatively in the
case $m^2 \ll \lt v^2$, as
\beqa
  \phi_m &=& v\lkk 1+ \frac{m^2}{\lt v^2}
  -\frac{3}{2}\lmk\frac{m^2}{\lt v^2}\rmk^2 + \cdots\rkk, \\
  V_0 &=& \frac{\lt}{16}v^4 +
  \frac{1}{2}m^2v^2+\frac{m^4}{\lt}+\cdots.
\eeqa
Another parameter, say $\lt$, can be fixed from the amplitude of
large-scale CMB fluctuations using the COBE data as before.  Hence we
are essentially left with  two free parameters, $v$ and $m$.
While $v$ mainly controls the speed of $\phi$ around the origin
and its fate, {\it i.e.}, to which minimum it falls, and $m$ mainly
governs the duration of new inflation, the entire dynamics is
determined by a complicated interplay of the three parameters.
For example, we cannot determine $\lt$ until we calculate the
duration of new inflation which also depends on $\lt$ itself for
fixed values of $m$ and $v$.
Hence we must numerically solve the equations
of motion iteratively to find out
appropriate values of parameters to produce PBHs at the right scale
with the right amount.

\begin{figure}[htb]
  \begin{center}
%    \epsfile{file=potential.eps,width=10cm}
  \leavevmode\psfig{figure=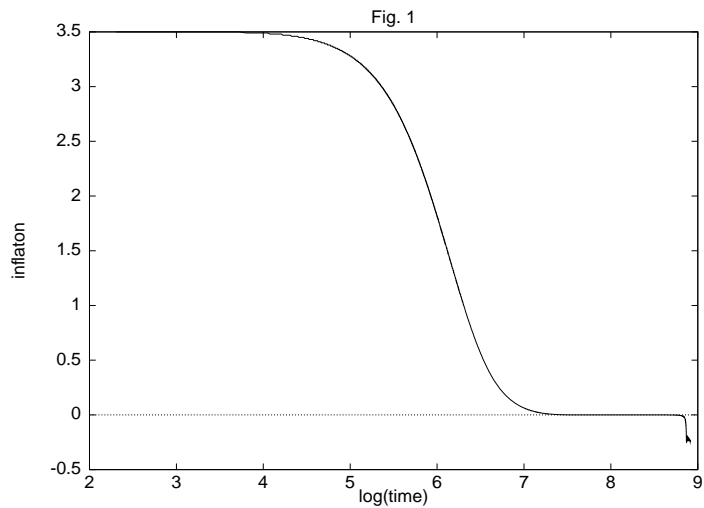,width=9.5cm}
  \end{center}
  \caption{Evolution of the inflaton in chaotic new inflation
with $\lt=3\times10^{-12}$, $m=6\times 10^{-8}\mpl$, and
$v=0.2138436\mpl$.  Time and $\phi$ are
 displayed in unit of the Planck time and $\mpl$,
respectively.}
  \label{fig:1}
\end{figure}

Let us now consider a specific example of formation of MACHO-PBHs.
For this purpose we must realize a peak with $\beta\sim 10^{-10}$ on the
comoving Hubble scale at $N=35$.
After some iterative trials we
have chosen $\lt =3\times 10^{-12}$ and $m=6\times 10^{-8}\mpl$,
and then solved the equation of motion for various values of $v$.
In this choice of $\lt$ and $m$ we find new inflation lasts for
more-than ten $e$-folds expansion if we take $v$ in the range
$v=0.2131\mpl-0.2147\mpl$.  Hence we do not need much fine-tuning
of the model parameters to realize a new inflationary stage
itself.  We also find that $\phi$ settles down to $\phi_m$ classically if
$v \geq 0.213843638\mpl$ and to $-\phi_m$ if $v \leq
0.213843631\mpl$.  If, on the other hand, we choose $v$ in the
range $0.213843632\mpl < v < 0.213843637\mpl$, $\phi$ spends more
than one expansion time in the region
$|\phi|<\phi_q=3.4\times 10^{-8}\mpl$ and the universe would enter
the self-reproduction regime.

\begin{figure}[htb]
  \begin{center}
%    \epsfile{file=potential.eps,width=10cm}
  \leavevmode\psfig{figure=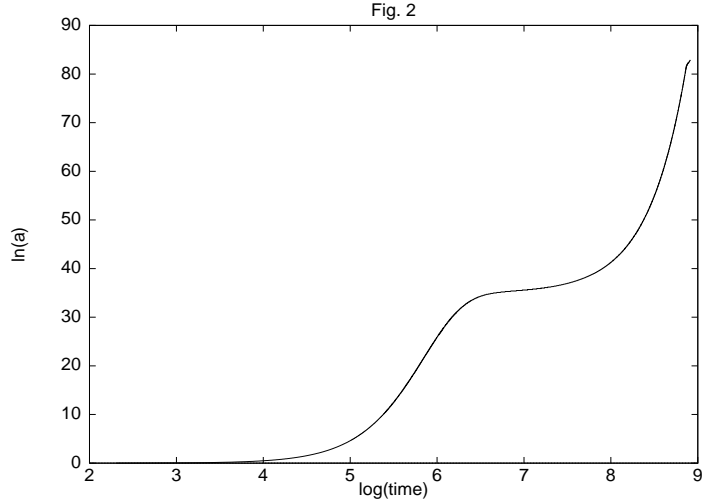,width=9.5cm}
  \end{center}
  \caption{Evolution of the scale factor with the same parameters.}
  \label{fig:2}
\end{figure}

Figures 1 and 2 depict evolution of the scale factor and the
inflaton $\phi$, respectively, for the case $v=0.21384360\mpl\equiv v_M$
with the initial condition $a_i=1$ at $\phi_i=3.5\mpl$.
The chaotic inflation ends at $\phi\simeq 0.89\mpl$ and new
inflationary expansion sets in at $\phi\simeq 0.082\mpl$ but
the slow roll-over phase starts only at
$\phi_s=-4.03\times10^{-7}\mpl\equiv \phi_M$.  In this case $\phi$
is found to stay in the region $|\phi|<\phi_q$ for only about 0.1 expansion
time.
The linear perturbation $\Delta$ has the right amplitude on large
scales to meet the COBE observation, and it has a peak on the
comoving Hubble scale at $\phi=\phi_s$.  We find
$N(\phi_s\rightarrow\phi_f)=35$, which is the right scale for
MACHO-PBHs.

In this case we find $\gamma=0.300590$ and the abundance of the
PBHs at formation reads
\beq
  \beta=2.29c\exp(-0.0196994c^{-2}), \label{49}
\eeq
with $h_{bh}=0.75$.
For $\phi_s=\phi_M$ we find the peak abundance of PBH,
$\beta=6\times 10^{-10}$ at the mass scale $M\simeq 1\msolar$\footnote{
In obtaining (\ref{49}) we have started with
$\Pi[\phi,t=t_s]=\delta(\phi-\phi_s)$ in order to extract information on
a specific mass scale.  This also corresponds to
treating that $\phi$ evolves along the classical trajectory until
$\phi=\phi_s$, which is correct only as an average.   In fact,
due to quantum fluctuations generated during chaotic
inflation and the early stage of new inflation, $\phi_s$ itself
takes different values in different domains.
We have confirmed, however, that $\phi_s$ shifts only
 about $\pm 4H_0/(2\pi)$ even if we consider their effects.
Since we find $|\phi_M| \gg 4H_0/(2\pi)$, it does not induce significant
fluctuations in $\beta$.  For the same reason we are free from
the domain wall problem which could be present if $\phi$ had a large
fluctuation and different regions fell different minima.}.
Using (\ref{cgam}) we can also write it as
\beq
  \beta=0.888\Delta_s\exp(-0.131072\Delta_s^{-2}).  \label{peakb}
\eeq

One can also obtain an approximate shape of the mass spectrum of
the PBHs using (\ref{peakb}) with $\Delta_s$ replaced by $\Delta$
at different epoch corresponding to different black hole mass.
More specifically the mass of black holes, $M$, and their initial
fraction, $\beta(M)$, can be written by an implicit function of $\phi$
as
\beqa
  M&=&K\exp\lmk 2\lkk N^{}(\phi\rightarrow\phi_f)-35\rkk\rmk\msolar,
  \label{mass}\\
  \beta(M)&\simeq&
  0.888\Delta(\phi)\exp\lmk -0.131072\Delta^{-2}(\phi)\rmk,
  \label{betad}
\eeqa
where $K$ is a factor of order of unity which depends on the
expansion law of the post-inflationary universe but we put $K=1$
for simplicity below.

Figure 3 depicts the mass spectrum of black holes obtained from
(\ref{mass}) and (\ref{betad}).  Thus the PBH abundance is sharply
peaked.  Note, however, that the shape of
the large-mass tail is not exactly correct which corresponds to
the regime where slow-roll solution is invalid.  Nonetheless this
figure correctly describes the location of the peak.

\begin{figure}[htb]
  \begin{center}
%    \epsfile{file=potential.eps,width=10cm}
  \leavevmode\psfig{figure=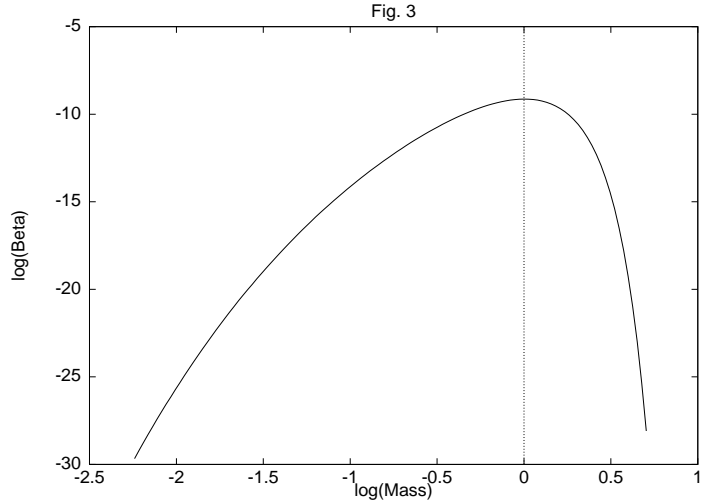,width=9.5cm}
  \end{center}
  \caption{Expected mass spectrum of PBHs in chaotic new
inflation with the same model parameters. Mass is displayed in
unit of the solar mass.}
  \label{fig:3}
\end{figure}

Apart from the effects of inflaton's detailed dynamics, however,
we must say that the above spectrum is only qualitatively correct,
because we have chosen a specific value of the threshold,
$h_{bh}=0.75$, and assumed the black hole mass is equal to the
horizon mass when the perturbed region reentered the Hubble
radius (see also \cite{NJ,41}).
In order to improve the calculation of the mass spectrum
we must calculate the probability distribution functional of the
configuration of the perturbed region and then calculate the final
mass of the black hole, if formed, for each configuration, which
is beyond the scope of the present analysis.  Nevertheless from
the above analysis we can
convince ourselves that we can produce large enough amplitude of
curvature fluctuation on a desired scale in the chaotic new
inflation scenario.

Within the limit of our predictability of the mass spectrum,
we can also apply our model for the formation of PBHs with
different masses and abundance.  For example, we may produce PBHs
with $M\simeq 10^8\msolar$ which may act as a central engine of
AGNs with the current density, say, $n\sim 10^{-5}{\rm Mpc}^{-3}$
corresponding to $\beta\sim 10^{-11}$ at formation.  From (\ref{mass}) we
find $M=10^8\msolar$ corresponds to $N(\phi_s\rightarrow\phi_f)=44$
and  the desired spectrum is realized for $\lambda=3.7\times 10^{-12}$,
$m=6.1\times 10^{-8}\mpl$,
and $v=0.21532324\mpl$ under the COBE normalization.

Another interesting possibility is to produce a tiny amount of
PBHs which are evaporating right now, with the initial mass
$M\simeq 10^{15}$g.  With the current abundance $\Omega\simeq 10^{-8}$
or $\beta\simeq 10^{-25}$ at formation, they may explain a class
of gamma-ray bursts.  In this case we should have only a short
period of slow-roll new inflation,
$N(\phi_s\rightarrow\phi_f)=14$.
We find $\beta=2\times 10^{-25}$ at the right mass scale if we choose
 $\lambda=3\times 10^{-13}$,
$m=5\times 10^{-8}\mpl$, and $v=0.16557604828\mpl$.
In this case a relatively large value of $m$ is required in order
to keep new inflation short, and we cannot necessarily rely on the
stochastic inflation method which is valid only if $|m^2| \lsim
H^2_0$.  This does not mean we cannot generate large enough
density fluctuation on the relevant scale.  The only problem is
we do not have a reliable method to calculate magnitude of
fluctuation or black hole
abundance accurately in such situations.

\section{Conclusion}

In the present paper we proposed a new scenario of double inflation
which contains only one inflation-driving scalar field, that is, we
pointed out that a scalar potential which has an unstable
local maximum at the origin can not only realize new and chaotic
inflation separately but also accommodate both sequentially if its
model parameters are appropriately chosen with natural initial
conditions as employed in the chaotic inflation scenario.
We have further shown that the spectrum of density
fluctuation in this model can have a large-amplitude peak on the
comoving Hubble scale at the onset of the slow-roll regime of new
inflation and that this can be applied to formation of PBHs on a
specific mass scale. This feature of the spectrum is realized
naturally compared with other models with  a single scalar field
\cite{INN,BP} because the scalar field is not slowly rolling at
the onset of new inflation.  On the other hand, we must specify the
values of model parameters with many digits  in order to produce
an appropriate amount of PBHs on a desired scale.  This feature,
however, is more or less common to all the other models attempting
to account for formation of PBHs in inflationary cosmology (see {\it e.g.}
\cite{BP}),
because both the peak amplitude of the fluctuations and its
location must be specified with high accuracy due to the
exponential dependence of the black hole abundance and its mass on
the model parameters.

On the other hand, one could in principle claim that observation
of PBHs can serve as a strong tool to determine the parameters in
the inflaton's dynamics.  Unfortunately, however,  our
ignorance of the detailed condition for PBH formation, such as the
precise threshold amplitude of fluctuation as a functional of the
shape of perturbed region, makes it impossible to link the mass
spectrum of PBHs with the shape of the inflaton's potential precisely.
In the present paper we have calculated the values of the model
parameters rather precisely under the universal assumption of
$h_{bh}=0.75$.  In fact, however, the values of the parameters
would totally change had we chosen a different threshold.  Hence
the precise numbers we have quoted do not have much significance,
but the number of digits simply indicates the sensitivity of the mass
spectrum to the model parameters.

\acknowledgements{
The author is grateful to Professor Andrei Linde for useful comments and
his hospitality at Stanford University, where this work was done.
This work was partially supported by the Monbusho.
}


\begin{thebibliography}{99}
\bibitem{Zel}Ya.B.\ Zel'dovich and I.D.\ Novikov, Soviet Astronomy
{\bf 10}, 602 (1967).
\bibitem{Haw}S.W.\ Hawking, Mon.\ Not.\ R.\ astr.\ Soc. {\bf 152}, 75 (1971).
\bibitem{Carr}B.J.\ Carr, Astrophys.\ J. {\bf 201}, 1 (1975).
\bibitem{Nov}B.J.\ Carr, Astrophys.\ J. {\bf 206}, 8 (1976);
S.\ Miyama and K.\ Sato, Prog.\ Theor.\ Phys. {\bf 59},
 1012(1978);
I.D.\ Novikov, A.G.\ Polnarev, A.A.\ Starobinsky,
Ya.\ B.\ Zel'dovich, Astron.\ Astrophys. {\bf 80}, 104 (1979).
\bibitem{macho}C.\ Alcock et al., Nature, {\bf 365}, 623 (1990); Phys.\
 Rev.\ Lett. {\bf 74}, 2867 (1995); Astrophys.\ J. {\bf 486}, 697 (1997);
E.\ Aubourg et al., Nature, {\bf 365}, 623 (1993); Astron.\ Astrophys.
{\bf 301}, 1 (1995).
\bibitem{RF}H.B.\ Richer and G.G.\ Fahlman, Nature 358, 383 (1992).
\bibitem{BU}S.P.\ Boughn and J.M.\ Uson, Phys.\ Rev.\ Lett. {\bf 74},
216 (1995).
\bibitem{pop3}B.J.\ Carr, J.R.\ Bond, and W.D.\ Arnett, Astrophys.\ J.
{\bf 277}, 445 (1984).
\bibitem{Naka}T.\ Nakamura, M.\ Sasaki, T.\ Tanaka, and K.S.\ Thorne,
Astrophys.\ J.\ Lett. {\bf 487}, L139 (1997).
\bibitem{evaporate}S.W.\ Hawking, Nature {\bf 248}, 30 (1974);
Comm.\ Math.\ Phys. {\bf 43}, 199(1975).
\bibitem{Cline}D.\ Cline, D.A.\ Sanders, and W.\ Hong, Astrophys.\ J.
{\bf 486}, 169(1997).
\bibitem{Turner}E.L.\ Turner, Astron.\ J. {\bf 101}, 5(1991).
\bibitem{Loeb} A.\ Loeb, Astrophys.\ J. {\bf 403}, 542(1993);
M.\ Umemura, A.\ Loeb, and E.L.\ Turner, Astrophys.\ J. {\bf 419},
459(1993).
\bibitem{oriinf} A.H. Guth, { Phys.\ Rev.} {\bf D23}, 347 (1981);
 K. Sato, { Mon.\ Not.\ R.\ astr.\ Soc.} {\bf 195}, 467 (1981).
\bibitem{newinf} A.D.~Linde, Phys.\ Lett. {\bf 108B}, 389 (1982);
A.\ Albrecht and P.J.\ Steinhardt, Phys.\ Rev.\ Lett. {\bf 48}, 1220
(1982).
\bibitem{chaoinf} A.D.~Linde, Phys.\ Lett. {\bf 129B}, 177
(1983).
\bibitem{inf} For a review of inflation see, $e.g.$ A.D.\ Linde,
Particle Physics and Inflationary Cosmology (Harwood, 1990).
\bibitem{pert} S.W.\ Hawking, { Phys.\ Lett.} {\bf 115B}, 295 (1982);~
A.A.~Starobinsky, Phys.\ Lett. {\bf 117B}, 175 (1982);~
A.H. Guth and S-Y. Pi, { Phys.\ Rev.\ Lett.} {\bf 49}, 1110 (1982).
\bibitem{COBE}C.L.\ Bennet, {\it et al}. Astrophys.\ J.\ Lett. {\bf464},
 L1 (1996).
\bibitem{Hodges}H.M.\ Hodges and G.R.\ Blumenthal, Phys.\ Rev.
{\bf D42}, 3329 (1990).
\bibitem{INN}P.\ Ivanov, P.\ Naselsky, and I.\ Novikov, Phys.\
Rev. {\bf D50}, 7173 (1994).
\bibitem{BP}J.S.\ Bullock and J.R.\ Primack, Phys.\ Rev.
{\bf D55}, 7423 (1997).
\bibitem{JY}J.\ Yokoyama, Astron.\ Astrophys. {\bf 318},
673 (1997).
\bibitem{ST}J.\ Silk and M.S.\ Turner, Phys.\ Rev.\ {\bf D35},
419 (1987).
\bibitem{Ra}L.\ Randall, M.\ Soljaci\'c, and A.H.\ Guth, Nucl.\ Phys.
{\bf B472}, 377 (1996).
\bibitem{GL}J.\ Garc\'ia-Bellido, A.D.\ Linde, and D.\ Wands,
Phys.\ Rev. {\bf D54}, 6040 (1996).
\bibitem{Kawa}M.\ Kawasaki, N.\ Sugiyama, and T.\ Yanagida,
hep-ph/9710259.
\bibitem{di}
L.F.\ Kofman, A.D.\ Linde, and A.A.\ Starobinsky, Phys.\ Lett.\ B
{\bf 157}, 361 (1985); R.\ Holman, E.W.\ Kolb, S.L.\ Vadas,
and Y.\ Wang, Phys.\ Lett.\ B {\bf 269}, 252 (1991);
 D.~Polarski and A.A.~Starobinsky, Nucl.\ Phys. {\bf
B385}, 623 (1992); ~Phys.\ Rev. {\bf D51}, 6123 (1994);
S.~Gottl\"ober, J.P.\ Muecket, and
A.A.~Starobinsky, Astrophys.\ J. {\bf 434}, 417 (1994).
\bibitem{BK}R.H.\ Brandenberger and J.H.\ Kung, Phys.\ Rev. {\bf D42},
 1008 (1990).
\bibitem{topological}A.D.\ Linde, Phys.\ Lett.\ B {\bf 327}, 208 (1994);
A.\ Vilenkin, Phys.\ Rev.\ Lett. {\bf 72}, 3137 (1994).
\bibitem{CW}S.\ Coleman and E.\ Weinberg, Phys.\ Rev. {\bf D7}, 788
 (1973).
\bibitem{SW}R.K.\ Sachs and A.M.\ Wolfe, Astrophys.\ J. {\bf 147}, 73 (1967).
\bibitem{Salopek}D.S.\ Salopek, Phys.\ Rev.\ Lett. {\bf 69}, 3602 (1992).
\bibitem{eternal}A.\ Vilenkin, Phys.\ Rev. {\bf D27}, 2848 (1983);
A.D.\ Linde, Phys.\ Lett.\ B {\bf 175},
395 (1986); Mod\ Phys.\ Lett. {\bf A1}, 81 (1986).
\bibitem{Staro} A.A.\ Starobinsky, In Field Theory, Quantum Gravity, and
 Strings, eds.\ H.J.\ de Vega and N.\ Sanchez, Lecture Notes in Physics
Vol.\ 246 (Springer, Berlin, 1986), 107.
\bibitem{Ivanov}P.\ Ivanov,  Phys.\ Rev.\ {\bf D57}, 7145 (1998).
\bibitem{KS}H.\ Kodama and M.\ Sasaki, Prog.\ Theor.\ Phys.\ Suppl.
{\bf 78}, 1 (1984).
\bibitem{NNP}D.K.\ Nadezhin, I.D.\ Novikov, and A.G.\ Polnarev,
Sov.\ Astron. {\bf 22}, 129 (1978)
\bibitem{BH}G.V.\ Bicknell and R.N.\ Henriksen,
Astrophys.\ J. {\bf 232}, 670 (1978).
\bibitem{NJ} J.C.\ Niemeyer and K. Jedamzik, Phys.\ Rev.\ Lett.
{\bf 80}, 5481 (1998).
\bibitem{41}J.\ Yokoyama,  gr-qc/9804041 To be published in Physical Review D.
\end{thebibliography}
\end{document}